\documentstyle[epsf,12pt]{article}
\begin{document}

\newcommand{\etal}{{\it et al\,., }}
\def\sq{\mathord{\not\mathrel{q}}}
\def\sk{\mathord{\not\mathrel{k}}}
\def\s0{\mathord{\not\mathrel{0}}}
\def\se{\mathord{\not\mathrel{\epsilon}}}
\begin{flushright}
KEK CP-056 \\
KEK Preprint 97-68
\end{flushright}

\vspace*{4cm}
\begin{center}
{\bf \LARGE AUTOMATIC CALCULATION OF COMPLETE}\\ ~~\\
{\bf \LARGE $O(\alpha)$ CORRECTIONS\ TO 
               $e^+e^- \rightarrow W^+\mu \bar{\nu}_{\mu}$}\\

\vspace{4mm}

J. Fujimoto, T. Ishikawa, Y. Kurihara, Y. Shimizu,\\
High Energy Accelerator Research Organization(KEK) \\
Oho 1-1, Tsukuba, Ibaraki 305, Japan \\

\vspace{3mm}

K. Kato, N. Nakazawa, \\
Kogakuin University \\
Nishi-Shinjuku 1-24, Shinjuku, Tokyo 160, Japan \\

and  \\
  
T. Kaneko \\
Meiji-Gakuin University \\
Kamikurata 1518, Totsuka, Yokohama 244, Japan \\

\end{center}

\vspace{6mm}
\begin{center}
Presented by J. Fujimoto
\end{center}
\vspace{6mm}
\begin{abstract}
Using the automatic system {\tt GRACE-LOOP}, the full $O(\alpha)$ 
electroweak corrections has been calculated for the process 
$e^+e^- \rightarrow W^+\mu \bar{\nu}_{\mu}$. The total correction 
to the cross section is found to be typically $-6.4$\% at $\sqrt{s}=190$
GeV with $10^\circ$ cut on the muon angle from the beam, including the 
correction from the hard photon emission. It is observed that 
the correction is rather sensitive to the physical parameters.
With the same conditions the correction corresponding to the real 
$W$-boson pair is $-4.1$\% and hence the deviation is not negligible.
\end{abstract}
\vspace{1cm}
{\it To appear in 1997 Moriond proceedings on "Electroweak Interactions
and Unified Theories"}


\newpage
\section{Introduction}

To study the $W$-boson properties in LEP2 experiments  
one needs a generator which can produce 4-fermion final 
states, since one can observe only these states but not the $W$-boson 
pair directly. This brings some technical difficulties in 
the calculation; there are so many(76) 4-fermion processes 
induced by $e^+e^-$ collisions and the number of Feynman 
diagrams in each process is also big, for instance 144 for the case of 
$e^+e^-e^+e^-$ even in the tree level.
In performing such 
a large scale computation it is desirable to have an automatic 
calculating system. 
Such a system {\tt GRACE}\cite{GRACE} has been developed 
for the tree level calculation and utilized 
to create a generator in the most efficient way\cite{grc4f}.
 
Including the loop diagrams several works\cite{ww} have shown that the 
$O(\alpha)$ full electroweak corrections to the on-shell $W$-pair 
production become sizeable  depending on the physical parameters, such 
as the mass of top-quark.  Then it is interesting 
to see what is the $O(\alpha)$ corrections to the 4-fermion state and 
to compare them with those of $W$-pair. This is a study also 
indispensable for a precise data analysis. Till now, unfortunately, 
there exists {\sl no} complete calculation of the corrections except for
those from the hard photon emission or from a very limited number of 
loops\cite{LEP2}.
The reason hard to carry out the full 
calculation is clearly a huge number of Feynman diagrams involved. 
For instance there are about 1,000 diagrams for 
$e^+e^- \rightarrow \mu\bar{\nu}_{\mu}{u}\bar d$, and it amounts more 
than 9,000 for the case of two $e^+e^-$ pairs.

As an intermediate step and also as an exercise toward various 4-fermion
processes, we report here the full $O(\alpha)$ corrections to 
$e^+e^- \to W^+\mu\bar\nu_{\mu}$. This is unsatisfactory in regard with
the final goal, the 4-fermion case, but certainly goes one step further 
from the real $W$-pair approximation.
   
\section{One-loop Corrections to 
              $e^+e^- \rightarrow W^+\mu\bar\nu_{\mu}$}

\def\bar{\overline}

There are 6 tree diagrams and 287 loops which contribute to the 
$O(\alpha)$ corrections. Feynman diagrams and corresponding matrix 
elements were generated by the automatic calculation system 
{\tt GRACE-LOOP}\cite{gracel1}. In addition the radiative process 
$e^+e^- \rightarrow W^+\mu\bar\nu_{\mu}\gamma$ with a hard 
photon($k>k_c$) has been calculated with {\tt GRACE}, which is also
necessary to complete the total correction. We assume the Feynman gauge 
throughout the calculation of the one-loop diagrams.

First we give the result of this work. Some technical aspects will
be described later. In the estimation we have restricted the scattering 
angle of the final muon to $\vert\cos\theta_{\mu}\vert<\cos 10^{\circ}$,
to avoid a numerical instability which happened in a few loop diagrams. 
The unpolarized total cross section with full $O(\alpha)$ corrections
is then 1.720$\pm$0.008 pb at $\sqrt{s}=190$GeV for the following 
physical parameters: $M_Z=91.187$GeV, $M_W=80.37$GeV, $\Gamma_W=2.05$GeV,
$M_{Higgs}=300$GeV and $m_{top}=180$GeV. Here $-1.413$ pb comes
from the one-loop diagrams and the soft radiation with the photon 
energy cut $k_c=100$ MeV, and 1.295 pb from the hard radiation. 
The corresponding tree cross section is 1.838 pb. Hence the total 
radiative correction to the lowest order is $-6.4\pm0.4$ \%.
Figure 1(a) and (b) depict the scattering angle distribution 
of $W^+$ and $\mu^-$, respectively. Angles $\theta_W$ and
$\theta_\mu$ are defined from the initial electron. The histogram 
represents the distribution of the tree process and the black
squares indicate that corrected to $O(\alpha)$. The deviation seen
around the forward muon angle( backward $W^{+}$) is similar to that for
$W$-pair, which is due to the hard radiation.

Let us compare this result with the case of the real $W$-boson 
pair production, $e^+e^-\to W^+W^-$. This process has 3 tree and 140 
one-loop diagrams. With the same set of physical parameters but 
without any cut on $W^\pm$ we found the total correction of
$-4.1$\% with the tree cross section of 18.04 pb. 
It is interesting that
the corrections are sensitive to the top-quark mass, which appears 
in the quantities relevant to the two-point functions. This is common 
to both processes; for example $-1.4$ \%($W\mu\nu$) and almost 
0\%($W$-pair) for $M_W=80.22$ GeV and $m_{top}=130$ GeV.

One can see that the difference between these two processes is not 
negligible. Several sources could contribute to this. For $W\mu\nu$ 
there remain 147 diagrams that cannot be reduced to the double 
$W$-resonant process. Off-shellness of $W^-$ also gives some portion. 
A detailed analysis of the difference is now under investigation.
\begin{figure*}[t]
\centerline{\epsfbox{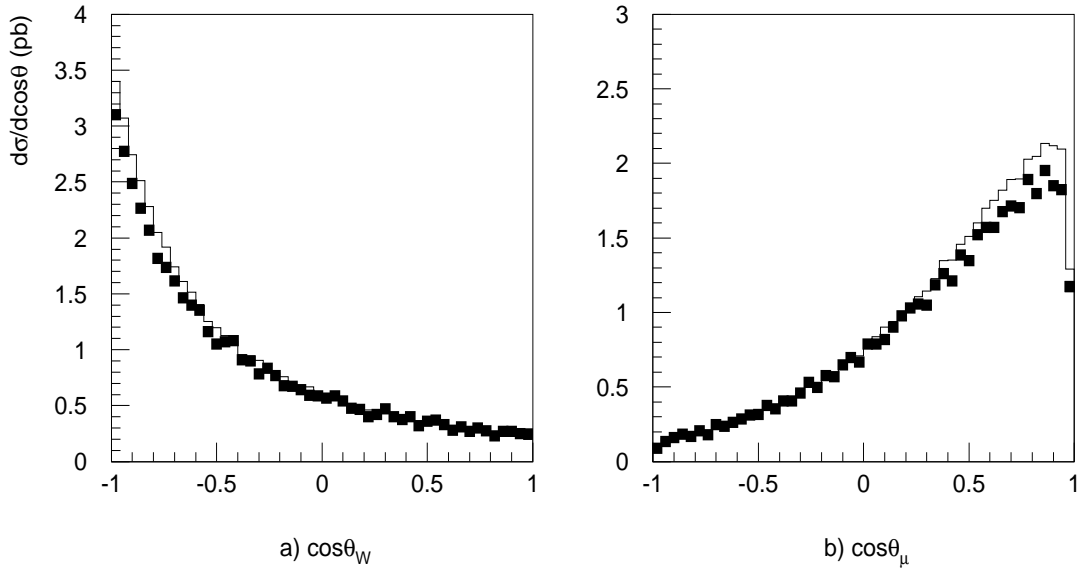}}
\caption{Scattering angle distribution of a) $W^+$ and of b) $\mu$.}
\end{figure*}

Next we briefly touch the technicality of the calculation, particularly 
how it was checked. First of all the system has been successfully 
applied to several $2 \to 2$ processes to reproduce the known 
results\cite{gracel1}. Hence we are convinced that every part of 
the system works correctly, {\sl i.e.}, the amplitude generation, 
the loop integrations for 3- and 4-point functions, the renormalization 
constants and the renormalized self-energies. 

However, since the process interested here is 2 to 3-body, 
the following new aspects appear which are absent in any 2 to 
2-body process: The multi-dimensional integration in the phase 
space and an evaluation of 5-point functions.
The 3-body kinematics is already equipped in the library of the 
automatic system. The Feynman parameters are numerically integrated 
by a Monte Carlo method together with the kinematical variables in the
phase space\cite{loopint}. We would like to emphasize that this way of
integration provides enough accurate results, with an uncertainty of 
less than 0.1\%. The 5-point function is estimated by a known reduction 
formula which relates the former to a sum of five 4-point 
functions\cite{5to4}.

We have performed the following checks as the evident self-test of 
the calculation:
(1) Renormalization.
\(1/\epsilon\) from UV-divergence is kept as a variable
in the program. The result was independent of this variable, as it 
should be.
(2) Infra-red divergence. 
This divergence appearing in some loop diagrams is regularized 
by a fictitious photon mass \(\lambda\). 
This is also kept in the program and it was checked whether 
the dependence is compensated by the soft 
radiation($k<k_c$)\cite{gracel1}.
(3) Soft-photon cut independence. 
Independence of the unphysical soft photon energy cut parameter $k_c$ 
was observed, which separates the soft radiation(analytic formula) and 
the hard one(Monte Carlo).

\section{Summary }

By using {\tt GRACE-LOOP}, designed to automate the evaluation of 
the one-loop diagrams in the standard model, a numerical estimation has 
been carried out, for the first time, to get the full $O(\alpha)$ 
corrections to $e^+e^- \to W^+\mu \bar{\nu}_{\mu}$. This process is 
still not a realistic one because the $W^+$ is treated as a
real particle. Although our estimation has such a limitation, we found 
that the total correction shows a deviation from that of $W$-pair case. 
The difference is not small to be ignored completely, though these
two corrections are of course comparable in magnitude. 
From the view point of the precise measurement it is very interesting 
and urgent to look how much the corrections are for various 4-fermion 
final states, such as $e^+e^- \rightarrow \mu\bar{\nu}_{\mu}{u}\bar d$.

\vspace{3mm}
\noindent {\bf Acknowledgments }

This work has been done as a part of the {\tt GRACE} project under 
the KEK-LAPP collaboration. We would like to thank M. Kuroda and
colleagues at LAPP, particularly, Drs. P. Aurenche, F. Boudjema, 
G. Coignet, J.-Ph. Guillet and D. Perret-Gallix for their thorough 
interest and 
encouragement. One of the authors(J.F.) is grateful to Prof. Tr\^an 
Thanh V\^an for providing him with an opportunity to present the 
current work at the Electroweak session of Moriond '97. 
This work was supported in part by the Grant-in-Aid(No.07044097) of 
Monbu-sho, Japan.

%
%
%
%
%
%
%
%
%

\begin{thebibliography}{99}

%
\bibitem{GRACE}
T. Kaneko \etal {\it New Computing Techniques in Physics
Research II}, ed. D. Perret-Gallix, World Scientific, 1992, 659;
T. Ishikawa \etal {\it GRACE manual}, KEK report 92-19 (1993). 
%
\bibitem{grc4f}
 J. Fujimoto \etal Comput. Phys. Commun. {\bf 100}(1996) 128.
%
\bibitem{ww}
M. Lemoine and M. Veltman, Nucl. Phys. {\bf B164}(1980) 445;
M. B\"ohm, A. Denner, T. Sack, W. Beenakker, F. Berends,
H. Kuijf, Nucl Phys. {\bf B304}(1988) 463;
J. Fleischer, F. Jegerlehner and M. Zralek, Z. Phys. {\bf C42}(1989) 409;
J. Fleischer, F. Jegerlehner and K. Kolodziej, Phys. Rev. {\bf D47}
(1993) 830; M. Kuroda {\sl et al.}, in preparation.
%
%
\bibitem{LEP2}
Physics at LEP2, vol.1 and 2, CERN 96-01, Eds.
G. Altarelli, T. Sj\"ostrand and F. Zwirner; W. Beenakker 
{\sl et al.}, hep-ph/9612260; S. Dittmaier, Acta Physica Polonica
{\bf B28}(1997) 619; W. Beenakker {\sl et al.}, CERN-TH/97-114.

%
\bibitem{gracel1}
J. Fujimoto \etal Acta Physica Polonica {\bf B28}(1997) 945;
J. Fujimoto \etal  
{\it Proceedings of X-th Workshop on HEP and QFT}, ed. 
B.B. Levtchenko, (1996).
%
\bibitem{loopint}
J. Fujimoto, Y. Shimizu, K.  Kato and Y. Oyanagi,
Progr. Theor. Phys. {\bf 87} (1992) 1233.
\bibitem{5to4}
D.B. Melose, Nuovo Ciment {\bf XL A No. 1}(1965) 181,
W.L. van Neerven and J.A.M. Vermaseren, Phys. Lett. {\bf 137B}(1984) 
241;
A. Denner, Fortshr. Phys. {\bf 41}(1993) 307.
%
%
\end{thebibliography}
\end{document}